\documentclass[twocolumn,prl,aps,showpacs]{revtex4}
\begin{document}
\newcommand{\nd}{\noindent}
\newcommand{\be}{\begin{equation}}
\newcommand{\ee}{\end{equation}}
\newcommand{\ben}{\begin{eqnarray}}
\newcommand{\een}{\end{eqnarray}}
\newcommand{\nn}{\nonumber \\}
\newcommand{\ii}{\'{\i}}
\newcommand{\pp}{\prime}
\newcommand{\expq}{e_q}
\newcommand{\lnq}{\ln_q}
\newcommand{\quno}{q-1}
\newcommand{\qunoinv}{\frac{1}{q-1}}
\newcommand{\tr}{{\mathrm{Tr}}}

\draft

\title{Reciprocity relations between ordinary temperature and the
Frieden-Soffer's Fisher-temperature}
\author{F.~Pennini}

\author{A.~Plastino}

 \address{Instituto de F\'{\i}sica La Plata (IFLP)\\
 Universidad Nacional de La Plata (UNLP) and Argentine National
 Research Council (CONICET)\\ C.C.~727, 1900 La Plata, Argentina}


\begin{abstract}
Frieden and Soffer conjectured some years ago the existence of a
``Fisher temperature" $T_F$ that would play, with regards to
Fisher's information measure $I$, the same role that the ordinary
temperature $T$ plays vis-a-vis Shannon's logarithmic measure.
Here we exhibit the existence of reciprocity relations between
$T_F$ and $T$ and provide an interpretation with reference to the
meaning of $T_F$ for the canonical ensemble. \vskip 2mm


\end{abstract}
 \maketitle

\section{Introduction}

Frieden and Soffer conjectured some years ago
\cite{Frieden,roybook} the existence of a ``Fisher temperature"
$T_F$ that would play, with regards to Fisher's information
measure $I$, the same role that the ordinary temperature $T$ plays
vis-a-vis Shannon's logarithmic measure $S$ \cite{katz,jaynes}. In
a series of more recent publications, this conjecture was amply
validated by showing that the Legendre transform structure of
thermodynamics can be replicated  without changes if ones
substitutes $I$ for the  Jaynes-Shannon entropy $S$~\cite{pp1,pp2,pp3,pp4},
 which yields then  a ``Fisher
thermodynamics". A question still lingers, though: what is the
relation between $T$ and $T_F$? In this note we purport to provide
a first answer in this respect.

\section{Brief Fisher considerations}

Estimation theory \cite{roybook} provides one with a powerful
result with reference to  a system that is specified by a physical
parameter $\theta$. Let {\bf x} be a stochastic variable and
$p_\theta({\bf x})$ the probability density for this variable,
which depends on the parameter $\theta$.  If an observer
\begin{itemize}
\item  makes a measurement of
 ${\bf x}$ and
wishes  to best infer $\theta$ from this  measurement,
 calling the
  resulting estimate $\tilde \theta=\tilde \theta({\bf x})$ and
  \item wonders how well $\theta$ can be determined, \end{itemize}
  \nd then estimation theory
 asserts \cite{roybook} that the best possible estimator $\tilde
 \theta({\bf x})$, after a very large number of ${\bf x}$-samples
is examined, suffers a mean-square error $e^2$ from $\theta$ that
obeys a relationship involving Fisher's $I$, namely, $Ie^2=1$,
where the Fisher information measure $I$ is of the form
\begin{equation}
I=\int \,d{\bf x}\,p_\theta({\bf x})\,\left\{\frac{\frac{\partial
p_\theta({\bf x})}{
\partial \theta}}{p_\theta({\bf x})}\right\}^2 =
\left\langle   \left[ \frac{1}{p_\theta} \frac{\partial p_\theta}{
\partial \theta} \right]^2 \right\rangle \label{ifisher}.
\end{equation}
\nd This ``best'' estimator is called the {\it efficient}
estimator. Any other estimator must have a larger mean-square
error. The only proviso to the above result is that all estimators
be unbiased, i.e., satisfy $ \langle \tilde \theta({\bf x})
\rangle=\,\theta \label{unbias}$. Thus, Fisher's information
measure has a lower bound, in the sense that, no matter what
parameter of the system  we choose to measure, $I$ has to be
larger or equal than the inverse of the mean-square error
associated with  the concomitant   experiment. This result, i.e.,

\be \label{rao} I\,e^2\,\ge \,1,
 \ee
 is referred to as the
Cramer-Rao (CR) bound, and constitutes a very powerful statistical
result \cite{roybook}.

\section{Formalism}

  We start by defining the well known density operator that
describes a system at equilibrium~\cite{katz,jaynes}

 \be
\hat{\rho}=\frac{1}{Z}\,e^{-\sum_{i=1}^M \chi_i
\hat{A}_i},\label{rho} \ee where the $\chi_i$ are Lagrangian
multipliers associated to the $M-$observables $\hat{A}_i$ and \be
\label{constr} \langle
A_i\rangle=\mathrm{Tr}\hat{\rho}\hat{A}_i\,\,\,\,\,(i=1,\ldots,M),
\ee  where the partition function $Z$ is given by
$Z(\chi_i)=\mathrm{Tr}\left( e^{-\sum_{i=1}^M \chi_i
\hat{A}_i}\right)~$\cite{pathria1993}. In our present
considerations we assume that these multipliers  {\it have already
been determined}.

 Following Mandelbrot~\cite{Mandelbrot,Incerteza,powerlaw} we
will {\it associate} the above Lagrange multipliers  to parameters
to be estimated via {\it Fisher considerations}. We write down now
the Fisher information measure used in such an estimation
procedure, here associated to the probability distribution
$\hat\rho$~\cite{roybook,Incerteza,powerlaw,PRE04PP,PLA04PP} \be
I=\left\langle 
\,\sum_{i=1}^M\, \Gamma_i\, \left(\frac{\partial
\ln{\hat\rho}}{\partial \chi_i}\right)^2\right\rangle, \label{F1}
\ee where the $\Gamma_i$ are suitable constants related to the
need of expressing  $I$ as a dimensionless quantity, as discussed
in~\cite{PRE04PP,PLA04PP}. After replacing (\ref{rho}) into
(\ref{F1}) we then find that $I$ is intimately connected to our
observables' fluctuations, as  pointed out long ago by Mandelbrot~
\cite{Mandelbrot,Uffink}

\be I=\sum_{i=1}^M\,\Gamma_i\,\left\langle \left(
\hat{A}_i-\langle \hat{A}_i\rangle\right)^2\right\rangle. \ee It
is well known (and straightforwardly verified)  that the
statistical fluctuations of an observable obey the relation~\cite{Mandelbrot,Uffink}

\be \left\langle\left( \hat{A}_i-\langle
\hat{A}_i\rangle\right)^2\right\rangle= -\frac{\partial\langle
\hat{A}_i\rangle}{\partial \chi_i}, \ee which allows us to recast
the Fisher measure in the fashion \be
I=-\sum_{i=1}^M\,\Gamma_i\,\frac{\partial\langle
\hat{A}_i\rangle}{\partial \chi_i}\label{F2}. \ee


\nd As stated above, thermodynamics' Legendre structure can be
neatly re-obtained if one uses  Fisher's information instead of
Boltzmann' entropy~\cite{pp1,pp2,pp3,pp4}. We will be dealing
with the same mean values  $\langle \hat{A}_i\rangle$ used above,
but different Lagrange multipliers will ensue.
 Let us
call the Fisher multipliers $\gamma_i$ and borrow from~\cite{pp1}
the well known thermodynamic relation that links information
measure, Lagrange multipliers (here the Fisher ones), and
expectation values \cite{katz,pp1} \be \gamma_i=\frac{\partial
I}{\partial \langle \hat{A}_i\rangle}.\label{gamma} \ee It is now
clear that, introducing the above result into (\ref{F2}), we  get
an expression for the Fisher multipliers $\gamma_i$ in terms of
the Shannon ones

\be \gamma_i=-\sum_{j=1}^M\,\Gamma_j\, \frac{\partial}{\partial
\langle\hat{A}_i\rangle}\,\frac{\partial\langle
\hat{A}_j\rangle}{\partial \chi_j},\label{sist} \ee a relation
which could be used to  determine them. It might seem at this
point natural to ask: what happens if we consider a canonical
distribution in which the Lagrange multipliers are the $\gamma_i$
instead of the $\chi_i$?  We discuss this question show below for
classical systems within the strictures of the canonical ensemble.
\section{Equipartition theorem}

In classical statistical mechanics there exists a useful general
result concerning the energy $E$ of a system expressed as a
function of $N$ generalized coordinates $q_i$ and momenta $p_i$.
The result holds in the case of the following (frequent)
occurrence \begin{enumerate} \item the energy splits additively
into the form $E=\epsilon_i(r_i) + E'(q_1,\ldots,p_N),$ where
$\epsilon_i(r_i)$ involves only the degree of freedom $i$ (the
variable $r_i$)  and the remaining part $E'$ does not depend on
$r_i$. \item the function $\epsilon_i(r_i)$ is quadratic in $r_i$.
\end{enumerate}
In these circumstances $\langle \epsilon_i \rangle = kT/2, $ with
$k$ the Boltzmann's constant and $T$ the temperature. This is the
equipartition theorem \cite{reif}. Its demonstration assumes that
the  thermal equilibrium Bolztmann-Gibbs  equilibrium probability
distribution  \be f=\frac{1}{Z}\,e^{-\beta E} \label{rhoc} \ee
where $\beta=1/kT \equiv \lambda/k$ is the Lagrange multiplier
associated with the the mean-energy constraint $\langle E\rangle=
\int d\tau f E $, with $d\tau$ the phase-space volume element.
Setting $\Gamma=1/k$  yields a dimensionless Fisher
 information measure (\ref{F2}) for the canonical ensemble
\be I=-\frac{1}{k}\,\frac{\partial \langle E \rangle}{\partial
\lambda}.\label{IC} \ee

\section{Reciprocity}

\nd Since we assume equipartition we immediately find~\cite{reif}
$E= (Nk) \lambda^{-1}$.
 We have then \be \frac{\partial \langle E \rangle}{\partial
\lambda}= - Nk \lambda^{-2}=-\frac{E^2}{kN},\ee entailing that the
Fisher multiplier $\gamma$ is \be \gamma=\frac{1}{kT_F}= -
(1/k)\frac{\partial}{\partial \langle E \rangle}\frac{\partial
\langle E \rangle}{\partial \lambda}=\frac{2}{k \lambda}. \ee
Since the multipliers are inverse temperatures  we obtain the
interesting relationship

\be  T_F = \frac{1}{2T}, \label{interes}\ee our main result here,
which, on reflection, should  not surprise anyone since it is a
well known fact that whenever $I$ grows Shannon's $S$ decreases,
and viceversa~\cite{roybook}. \vskip 3mm

\nd As stated above, we introduce now $\gamma= 2/\lambda$ as the
multiplier entering the canonical probability distribution $f$ in
(\ref{rhoc}) and repeat the above discussion. Now $E=Nk/\gamma$
and \be \frac{\partial \langle E \rangle}{\partial \gamma}= - Nk
\gamma^{-2}=-\frac{E^2}{kN}.\ee We ask ourselves what is now

\be  \gamma_2=-\frac{1}{k}\frac{\partial}{\partial \langle E
\rangle}\frac{\partial \langle E \rangle}{\partial \gamma}\,\,?\ee
\nd Obviously,  \be \gamma_2 =\frac{2}{k \gamma}=
\frac{2}{k}\,\frac{k\lambda}{2}= \lambda=\frac{1}{T}. \ee

\nd Here we encounter reciprocity! The ``Fisher multiplier"
$\gamma_2$ is the inverse Boltzmann-Gibbs-Shannon temperature that
verifies

\ben \label{recipro} \frac{1}{kT}&=& \beta=\frac{\partial
I(\gamma)}{\partial \langle E\rangle_{\gamma}} \cr
\frac{1}{kT_F}&=& \gamma =\frac{\partial I(\beta)}{\partial
\langle E\rangle_{\beta}},\een in self-explanatory notation.

\section{Conclusions}

We have in this note provided two results that we deem important
for the Fisher practitioners, namely \begin{itemize} \item
$T_F=1/2T$ \item the reciprocity relations (\ref{recipro}).
\end{itemize}


\begin{references}



 \bibitem{Frieden}  B.R.~ Frieden and B.H.~ Soffer, {\it Phys.~Rev.~E} {\bf
  52}, 2274 (1995).

  \bibitem{roybook}  B.R.~ Frieden, {\it Physics from Fisher information}
  (Cambridge University Press, Cambridge, England, 1998).

 \bibitem{katz} A.~Katz, {\it Principles of Statistical Mechanics, The
  information
  Theory Approach} (Freeman and Co., San Francisco, 1967).


\bibitem{jaynes}
 E.T.~Jaynes, in: {\it Statistical Physics},
 ed.~W.K.~Ford
 (Benjamin, NY, 1963), p.~181.

 \bibitem{pp1} B.R.~Frieden, A.~Plastino, A.R.~Plastino and H.~Soffer,
   {\it Phys. Rev. E} {\bf 60}, 48 (1999).

 \bibitem{pp2}  R.~Frieden, A.~Plastino, A.R.~Plastino, B.H.~Soffer,  {\it Phys. Rev. E}
   {\bf 66}, 046128 (2002).


  \bibitem{pp3}  R.~Frieden, A.~Plastino, A.R.~Plastino, B.H.~Soffer,
{\it Phys. Lett. A} {\bf 304}, 73 (2002).

\bibitem{pp4} S.~Flego, B.R.~Frieden, A.~Plastino, A.R.~Plastino, B.H~Soffer,
{\it  Phys. Rev. E} {\bf 68}, 016105 (2003).

 \bibitem{pathria1993}
 R.K.~Pathria, {\it Statistical Mechanics} (Pergamon Press,
  Exeter, 1993).


\bibitem{Mandelbrot} B.~Mandelbrot, {\it Ann. Math. Stat.} {\bf
33}, 1021 (1962); {\it IRE Trans. Inform. Theory} {\bf IT-2}, 190 (1956);
 {\it J.~ Math.~ Phys.} {\bf 5}, 164 (1964).

\bibitem{Incerteza} F.~ Pennini, A.~ Plastino, A.R.~ Plastino and M.~ Casas,
{\it Physics Letters A} {\bf 302}, 156 (2002).

\bibitem{powerlaw} F.~Pennini, A.~Plastino,  {\it Physica A} {\bf 334}, 132 (2004).

\bibitem{PRE04PP}  F.~Pennini, A.~Plastino,  {\it Phys. Rev. E} {\bf 69}, 057101 (2004).

\bibitem{PLA04PP}  F.~Pennini, A.~Plastino, {\it Phys. Lett. A} {\bf 326}, 20  (2004).


\bibitem{Uffink} J.~ Uffink and J.~van Lith, {\it Foundations of
Physics} {\bf 29}, 655 (1999).

\bibitem{reif} F.~ Reif, {\it Statistical and thermal physics}
(McGraw-Hill, NY, 1965).

\end{references}
 \end{document}